\documentstyle[a4,twoside]{article}
\makeatletter
\renewcommand{\@oddhead}{Irregular Linear Hamilton Systems 
 \hfill \thepage}
\renewcommand{\@evenhead}{\thepage \hfill Sergej A. Choro\v savin }
\renewcommand{\@oddfoot}{}
\renewcommand{\@evenfoot}{}
\makeatother

{\vspace*{2ex}\par {#1}\quad\it }{\medskip\par }

\newenvironment{Thm}[2]%
{\par\addvspace{\bigskipamount}{\bf #1#2}\it }%
{\par\addvspace{\bigskipamount} }

\newenvironment{Definition}[1]{\begin{Thm}{Definition}{#1}}{\end{Thm} }

\newenvironment{Theorem}[1]{\begin{Thm}{Theorem}{#1}}{\end{Thm} }

\newenvironment{Lemma}[1]{\begin{Thm}{Lemma}{#1}}{\end{Thm} }

\newenvironment{Observation}[1]{\begin{Thm}{Observation}{#1}}{\end{Thm} }

\newenvironment{Remark}[1]{\begin{Thm}{Remark}{#1}}{\end{Thm} }

\newenvironment{Proof}{\par\addvspace{\bigskipamount} {\sc Proof}}%
{\par\hspace*{\fill}$\Box$ \par\addvspace{\bigskipamount} }

\newcommand{\dfrac}[2]{{\displaystyle \frac{#1}{#2}}}

\begin{document}
\begin{center}
{\bf Linear Hamilton Systems without Regular Properties.
 Solving a Problem Stated by M.G. Krein.}
\bigskip

 {{\sc Sergej A. Choro\v savin } }
\bigskip

\bigskip

 Email: { sergius@pve.vsu.ru }

\end{center}

\begin{abstract}
 We construct linear Hamilton systems without usual dichotomy
 property.
 The Ljapunov spectra of these systems are unfamiliar and conflicting,
 the behaviour of trajectories is very complicated. 
 The paper's subject refers to some problems 
 of indefinite inner product methods
 in the stability theory of abstract dynamical equation solutions.
\end{abstract}

\section%
 { Introduction }

 We start briefly describing some basic facts about 
 finite-dimensional linear autonomous invertible dynamical systems, 
 that is mathematically, 
 finite-dimensional systems of linear
\footnote{ when we say ``linear'' we mean ``properly linear'',
 that is,``linear homogeneous'' } 
 ordinary differential equations 
 with constant coefficients which are such that each separate 
 system 
 can be written in matrix form as 

$$
 \frac{dx(t)}{dt} = Ax(t) \,.
$$

 So, when we say 
 ``trajectory of the dynamical system'', we mean 
 ``solution to the corresponding ODE system''.

 Note, given a trajectory and another trajectory
 ---we say ``displaced trajectory''---, 
 then,  
 as we treat {\it linear} systems,
 the displacement {\it of} the former trajectory
 ---or, deviation {\it from} the former trajectory---
 is also a solution to the corresponding ODE system.

 Now then,
 suppose one treats trajectories of a linear dynamical systems.
 Then a standard classification fact is:

\medskip

{\it
 Every trajectory 
$x(t)$
 has a decomposition which is of the form 
$$
 x(t)=x_1(t)+\cdots+x_N(t) \leqno(*)
$$
 where 
$x_1(t),\ldots,x_N(t)$
 are trajectories of the {\bf same } System,
 which have usual (standard )
 exponential-wise behaviour,
 as 
$t\to\pm\infty$ .
}

\medskip

 If the finite-dimensional linear autonomous invertible dynamical system is,
 in addition,
 Hamilton, 
 then we observe a property ( of the trajectories ) of such a system, 
 we name that property ``being split'', 
 which is of an especial interest, especially in the theory of 
 {\bf non-linear (!!)} dynamical systems
\footnote{ we will not going into details,
 how does stability of {\bf non-linear } system relates to the 
 property ``being split'', which is defined for a {\bf linear } one
 }, 
 and which can be expressed 
 as 

\medskip

 {\bf Splitting Theorem. }{\it
 Every trajectory 
$x(t)$
 has a decomposition which is of the form 
$$
 x(t)=x_-(t)+x_+(t)\,;
\qquad
 |x_\pm(t)|\leq P_\pm(t) 
\qquad
 \hbox{ \rm as } t\to\pm\infty
\leqno(**)\,
$$
 where 
$P_\pm$
 are some polynomials and 
$x_\pm$
 are suitable trajectories of the {\bf same } dynamical system
}.

\medskip

 It is while one treats any linear finite-dimensional
 Hamilton system.
 {\bf But what about infinite-dimensional systems?}

\medskip

 First of all, we need to explain what we mean by 
 ``linear autonomous dynamical system''.
\begin{Definition}%
{ of abstract linear autonomous invertible dynamical system.}

\noindent
 Let 
$L$
 be a linear space.

\noindent
 Let a one-parameter family of linear operators on 
$L$,

$$
 \{U_{t}\}_{t}  \qquad  ( t \in {\bf R} ) \,,
$$
 be such that 
$$
  U_{t-r}U_{r-s} = U_{t-s} \,, \qquad \mbox{\rm if } t,r,s \in  {\bf R} \qquad 
  \makebox[12ex][l]{ {\bf ( consistency relation ) } }
$$
$$
  U_{0} = I = \mbox{ identity operator on } L \,.
$$

\par\addvspace{\medskipamount}\noindent
 In this case the pair 
$L, \{U_{t}\}_{t}$
 is said to be an 
 {\bf abstract linear autonomous invertible dynamical system}
 and 
 {\em two}-parameter family 
$ U_{t-s}\,,\, t,s \in {\bf R} $
 is called a {\bf propagator}, alias {\bf evolution operator}.

\par\addvspace{\medskipamount}\noindent
 Given a
$ t_0 \in {\bf R} $
 and an 
$x_0 \in L$,
 we say that the one-parameter family
$$
 x(t) =  U_{t-t_0}x_0  \qquad (t \in {\bf R})
$$ 
 is the {\bf trajectory}. In such a case we say that 
$t_0, x_0$
 are {\bf initial data}.

\par\addvspace{\medskipamount}\noindent

 In this paper we will mostly treat the case where the expression 
$t,s \in {\bf R}$ 
 is replaced by 
$t,s \in {\bf Z}$,
 i.e.,
$t,s$
 are integers.
 In this case we say that the dynamics is {\bf discrete} and write 
 something like 
$N,M$
 or 
$n,m$
 instead of 
$t,s$.
 Note that  
$$
 U_{N}= {\cal U}^{N} \qquad (\mbox{\rm for all integers } N) 
$$
 where 
${\cal U} := U_1$.
\par\addvspace{\medskipamount}\noindent

\end{Definition}

 In this paper we will treat only the case where the underlying space 
$L$  
 is {\bf Hilbert} (complex or real)
 and mostly the case where the dynamics is 
 {\em discrete}
\footnote{ some more details will be expounded later.
 As for a possible (and quite standard) generalization of the concept 
 of linear dynamics, see e.g. the section Appendix A. }.
 So, we need not immediately give
 any generalization of the concepts defined above.
 On the contrary,
 we shall restrict ourselves by special classes of propagators.

\medskip

 As already noted, we are interested in linear {\em Hamilton} systems.
 We adopt two different ways to formulate it in terms of propagator:

\par\addvspace{\medskipamount}\noindent
 1)\hfill 
\parbox[t]{0.95\textwidth}{
 the underlying space is symplectic (real or complex) and every 
$U_{t}$ 
 is a symplectic automorphism;

} 

\medskip

\noindent 
 2)\hfill 
\parbox[t]{0.95\textwidth}{
 the underlying space is $J$-space (complex or real) and every 
$U_{t}$ 
 is $J$-unitary.

} 

\par\addvspace{\bigskipamount}\noindent
 Commonly, the phrase 
 ``a linear operator, call it $T$, is {\bf $J$-unitary}''
 means:
$$
 T^*JT = J = TJT^*;\quad J^*=J;\quad J^2=I,
$$
 whereas if 
$T$
 and a
${\cal J}$
 are such that 
$$
 T^*{\cal J}T = {\cal J} = T{\cal J}T^*;\quad
 {\cal J}^*=-{\cal J};\quad {\cal J}^2=-I,
$$
 then 
${\cal J}$
 is an instance of {\bf operator of a symplectic structure} and 
$T$
 is a {\bf symplectic automorphism} 
\footnote{ linear symplectic automorphism is much more abstract object 
 than that the
$T$
 which satisfies 
$T^*{\cal J}T = {\cal J} = T{\cal J}T^*$
 for a 
${\cal J}$
 such that 
${\cal J}^*=-{\cal J};\quad {\cal J}^2=-I$ 
 } 
, which is also called
 linear canonical transformation and Bogoliubov transformation.
 We say ``{\bf $J$-unitary and symplectic automorphism}'' 
 instead of 
 ``$J$-unitary operator which is at the same time a symplectic automorphism''. 
 \qquad 

\bigskip

\underline{ First of all, it is necessesary to recall}

\begin{Lemma}{ H1-1}
 Let 
$V$
 be a linear bounded operator acting on a Hilbert space 
$H$.
 Suppose,
$V^{-1}$
 exists and is bounded.

 Then 
$V\oplus V^{*-1}$
 (i.e. the Hilbert direct sum of 
$V$
 and 
$V^{*-1}$) 
 is
$J$-unitary with respect to 
$J$
 which is defined by
$$
 J\,:\,x\oplus y\,\mapsto\,y\oplus x\quad (x\,,y\, \in H\,) \qquad
$$ 

 If one set instead of $J$ the operator which acts by the rule
$$
 {\cal J}\,:\,x\oplus y\,\mapsto\,y\oplus -x\quad (x\,,y\, \in H\,) \qquad ,
$$
 then one obtains that 
${\cal J}^*=-{\cal J};{\cal J}^2=-I$ ,
 i.e. 
${\cal J}$ 
 is an operator of a symplectic structure;
 in this case 
$V\oplus V^{*-1}$
 is a symplectic automorphism
.

$\Box$
\end{Lemma}

 We will systematically exploit the construction 
$V\oplus V^{*-1}$. 
 In that cases, we will write 
${\hat V} := V\oplus V^{*-1}$.

\bigskip

 Of course, the formulation of the Property ``being split''
 must be revised.
 At least two formulations seem to be appropriate:


\par\addvspace{\bigskipamount}\noindent
 1)\hfill 
\parbox[t]{0.95\textwidth}{
 Instead of letting only polynomials be in
$(**)$,
 it is reasonable to allow 
 the functions of ``not too rapid increase''
 to be present there:
 we will admit ``subexponentially increasing'' functions;
 in this case we say about the (regular) splitting ( or regular separation )
 of trajectories 
 and discuss the corresponding {\bf Existence Problem} of such trajectories;

 We specify the notion of ``being split'' by

\begin{Definition}{ {\rm of a} regularly split trajectory. }

\noindent
 If a trajectory 
$x(t)$
 has a decomposition of the form 
$$
 x(t)=x_-(t)+x_+(t)
\leqno(***)\,
$$
 where 
$x_\pm$
 are trajectories of the {\bf same } dynamical system, 
 and where 
$x_\pm$
 are such that 
$$
 (\forall \lambda > 0)
\quad
 \|x_+(t)\| e^{-\lambda t} \to 0
 \hbox{ \rm as } t\to +\infty
$$
$$
 (\forall \lambda > 0)
\quad
 \|x_-(t)\| e^{+\lambda t} \to 0 
 \hbox{ \rm as } t\to -\infty \,,
$$
 then we say that 
$x(t)$
 is {\bf regularly decomposable} or {\bf regularly split} 
 or, for brief,
$x(t)$
 is {\bf regular}. 

\end{Definition}

} 

\par\addvspace{\medskipamount}\noindent
 2)\hfill 
\parbox[t]{0.95\textwidth}{
 instead of seeking for the set of subexponentially increasing trajectories,
 one prefers to seek, having ideas of spectral theory in mind, for
 that invariant subspaces of the propagator of the system, 
 on which the spectral radius of the propagator would be 
$\leq 1$;
 in this case one says about {\bf Problem of {\sc M.G. Krein}}.

\par\addvspace{\bigskipamount}\noindent

}

\par\addvspace{\bigskipamount}\noindent
 It is important to take into account,
 that not only the case of the continuous ``time''
 (i.e. 
$t\in {\bf R}$),
 but also the discrete ``time'' case 
 ( primarily, 
$t\in {\bf Z}$ )
 or some other abstract ``time'' cases
 are interesting:
 we mean now ``symplectic representations of semi-groups''.
 As for phase spaces, i.e. the spaces
 on which such representations act, we repeat that 
 we must consider not only real, but also complex spaces.
%

\medskip

 To specify the Problems, 
 introduce a suitable definition.

\begin{Definition}{ D1-1}
$$
 S_0(T)\,:=\,\{x\in H|\quad\|T^Nx\|\to 0 \hbox{ for }N\to +\infty \},
$$ 
$$
 S(T)\,:=\,\{x\in H|\, \exists C\geq 0 \, \forall N\geq 0 
 \quad\|T^Nx\|\leq C \},
$$ 
$$
 S_+(T)\,:=\,\{x\in H|\forall a>1 \, \exists C\geq 0 \, \forall N\geq 0 
 \quad\|T^Nx\|\leq Ca^N\},
$$ 
$$
 r(T):= \hbox{spectral radius of }T \,.
$$
\end{Definition}

\begin{Remark}{ R1-1}
 Given a linear bounded operator 
$T$
 and a 
$T$-invariant subspace
$L$
 such that 
$r(T|L)\leq c$, 
 then 
$$
 L\subset S_+(c^{-1}T)
$$
\end{Remark}

\begin{Remark}{ R1-2}

$$
 S_{\rm x}(T_1\oplus T_2)=S_{\rm x}(T_1)\oplus S_{\rm x}(T_2);
$$
 here 
$S_{\rm x}$
 stands for 
$S_0$
 or 
$S$
 or 
$S_+$
 respectively.

\end{Remark}

\begin{Remark}{ R1-3}

$$
 S_0(T) \perp S(T^{*-1})
 \,,\qquad
 S(T) \perp S_0(T^{*-1}) \,,
$$
 which is immediate if one notices that 
$$
 |(x,y)|
 = 
 |(T^Nx,T^{*-N}y)|
 \leq
 \|T^Nx\|\,\|T^{*-N}y\| \,.
$$

\end{Remark}

 So, the first Problem is a problem of describing
 the structure of the set of the kind
$S_{\rm x}(T)$ .
 At least, one asks:
\begin{Thm}{Question}{ 1.}

 Let
$T$ 
 be any symplectic or 
$J$-unitary.

 Is 
$S_+(T) \not= \{0\}$ ?
\end{Thm}

 The second Problem looks more traditional: 
 one seek for the
$T$-invariant
 subspaces, 
 on which the previously given bounds of spectral radius are fulfilled. 
 At least, one asks:
\begin{Thm}{Question}{ 2.}

 Let
$T$ 
 be any symplectic automorphism or 
$J$-unitary operator.

 Is there a non-trivial 
$T$-invariant subspace 
$L$,
 such that 
$r(T|L) \leq 1$ ?
\end{Thm}

\par\addvspace{\bigskipamount}\par\noindent

 Perhaps it is just the time and place to recall
 the notion of Ljapunov indices.

\par\addvspace{\bigskipamount}\par\noindent
 Let 
$0$
 stand for the zero-element of the underlying space,
$x_0$
 be an element of the same space.
 We will refer to 
$x_0$
 as an instance of {\bf initial displacement} of 
$0$
 (or, as an instance of {\bf initial deviation} from 
$0$).

\noindent
 Let 
$t_0 \in {\bf R}$.
 We will refer to 
$t_0$
 as an {\bf initial value of the time},
$t$.

\noindent
 Let 
$\{x(t)\}_{t \in {\bf R}}$
 stand for the trajectory such that 
$x(t_0) = x_0$.
 Thus, 
$\{x(t)\}_{t \in {\bf R}}$
 is the displacement 
 of the zero trajectory starting at
$t=t_0$.
\footnote{ Notice, we now handle {\bf linear} systems }
\par\addvspace{\bigskipamount}\par\noindent
 The {\bf Ljapunov} upper {\bf indices}
 of the growth of the initial displacement,
$x_0$,  
 are real numbers, 
$\lambda_{\pm}$, 
 defined by
$$
 \lambda_{\pm}
  :=\limsup_{t\to \pm\infty}
            \frac{\ln \|x(t)\|}{|t|}
  =\inf\{
  \lambda|\: \|x(t)\|\leq C_{\lambda}e^{\lambda |t|}\, (t\to\pm\infty)
  \}
 \,,
$$ 
 Note, since the systems we handle are autonomous,
 neither 
$\lambda_{+}$
 nor 
$\lambda_{-}$
 depend on 
$t_0$.

\par\addvspace{\bigskipamount}\par\noindent

 So, the first Problem and Question 1, stated above, concern the associated
 theory of Ljapunov spectrum, 
 which is sometimes called Floquet-spectrum.

\par\addvspace{\bigskipamount}\par\noindent

 We will not go into historical details and restrict ourselves 
 by a simple enumeration of papers and books we dealt elaborating 
 the theme.
 They are : [Will36], [Will37]
 [Kre65], [KL1], [KL2], [IK], [IKL], [DalKre], [DadKul], [Kul], [Ikr89], 
 [Bogn], [DR], [Maj], 

 [Bog], [BraRob], [Ber], [Rob], [MV], [Emch], [RS2], [RS3], [Oks], 
 [DadKul], [Kul], [Ikr89].
 [Ch81], [Ch83], [Ch83T], [Ch84], [ChDTh].

\medskip

 Although the Problem of describing regular split displacements 
 is well-known for many years, it is still open.
 Even the Questions 1 and 2 have been answeren (in negative) not long ago,
 by [Ch97], [Ch98].
 
\medskip

 The present paper is based just on [Ch97], [Ch98]. 

\medskip

 More precisely, we construct three 
 discrete linear dynamical Hamilton systems 
 (the associated operators are 
$\hat{U}, \hat{V}, \hat{W}$
 in Sections 2, 3)
 and briefly describe their continuous analogues 
 (Section 4).
 All that systems have very complicated behaviour.

 Naturally, it is hardly worthy to qualify that systems as
 completely chaotic.

 However their Ljapunov spectra seem completely exotic,
 and the coresponding spectral subspaces and the sets 
$S_0, S, S_+ $ 
 seem to be strange and surprising.
 From this point of view we would rather say, 
 that the  behaviour of the constructed systems 
 is {\bf pre-chaotic}.

 In outline the situation is this:

 The first system we have constructed is such that 
$S_+$ 
 contains only element: this element is the zero element, certainly. 
 Hence, all Ljapunov indices are strictly positive. 
 Moreover, we take a number 
$c>0$ 
 quite arbitrarily, and then we construct a system, 
 such that {\bf both} of the Ljapunov indices 
$\lambda_{\pm}$ 
 of {\bf every } non-zero displacement 
$\geq\ln(2+c)$.
 And at the same time, there are displacements such that their 
 {\bf lower } index of growth 
$ = -\ln(2+c)$.

 In the case of the second system the set 
$S_+$ 
 is ``rich'': 
 There are ``many'' displacements such that their 
 Ljapunov indices are strictly negative definite,
 all they 
 $< -\ln 2 - \ln c$, $c>1$. 
 Nevertheless the closure of that set contains 
 some displacements such that their indices are strictly positive, 
 furthermore, they all 
$\geq \ln 2 + \ln c$.  

 As for third system, it is such that 
$S_0$ 
 is non-vanishing 
 ( so, 
$ \{0\} \not= S_0 \subset S \subset S_+$
), 
 and a space to be seeked for exists, i.e.
 a {\bf maximal} invariant subspace
$L$
 exists such that 
$r(\hat{W}|L) = 1 $
 (hence, $L \subset S_+$). 
 Nevertheless they, 
$L$
 and 
$S_0$,
 are mutually orthogonal, and furthermore, 
$L\cap \overline{S} = \{0\}$.

 Moreover, 
$L$ =${S_0}^{\perp}$, even 
$L$ =${S}^{\perp}$. 
 Moreover, the spectral radius of the propagator restricted on 
$L^\perp \equiv \overline{S_0}$ 
 is equal to 
$2$ and 
 the spectrum of the restriction itself is a subset of 
$\{z| 1\leq |z| \leq 2\}$.

 So are the facts that concern the Question 1. 

 As for Question 2 in itself, 
 for many years there was a suspiction that the answer is positive
 in every case.
 This suspiction was suggested by results of various kinds, 
 as by particular existence theorems---
 original [Kre64], relatively recent [Shk99]---,
 as by theorems of the sort---
 " 
$S_0(T)$
 is a subset of the subspace that had been constructed by M.G.Krein 
 [Kre64], [Kre65]"
 (for details see [Ch89-1], [Ch89-2], [Ch96T], [Ch00]).

 In spite of that the answer is negative.

\par\addvspace{\bigskipamount}\noindent
 We are now going to the exposition proper of the theme. 
 Troughout the paper, when we use ``$J$-Terminologie ''
 we have [Kre65] in mind.
 As for general mathematical terminology, we follow [RS].


\newpage\section
{ Discrete Dynamical System without Regular Trajectories }

\begin{Theorem}{ Th2-1}
 Let 
$c$
 be any real such that 
$c>0$.
 Then there exists a 
$J$-unitary and symplectic automorphism 
$\cal U$
 such that
$$
 S_+(c^{-1}{\cal U}) = S_+(c^{-1}{\cal U}^{-1}) = \{0\}\,.
$$

 In particular,

\par\addvspace{\medskipamount}\noindent 
{\rm (i)}\hfill 
\parbox[t]{.93\textwidth}{ 
 Let 
$L$
 be a 
$\cal U$-invariant subspace such that 
$L \not= \{ 0 \}$. 
 Then \mbox{$r({\cal U}|L) > c$; }
}
\par\addvspace{\smallskipamount}\noindent   
{\rm (ii)}\hfill 
\parbox[t]{.93\textwidth}{ 
 Let 
$L'$
 be a 
$\cal U$-invariant subspace such that 
$L' \not= \{ 0 \}$. 
 Then \mbox{$r({\cal U}^{-1}|L) > c$; }
}
\par\addvspace{\smallskipamount}\noindent  
{\rm (iii)}\hfill 
\parbox[t]{.93\textwidth}{ 
 there is no 
$\cal U$-invariant subspace 
$L''$
 such that 
 \mbox{$|spectrum\,({\cal U}|L'')|\geq c^{-1}$ } .
}
\par\addvspace{\medskipamount}\noindent 


\end{Theorem}

\begin{Proof}{ }

 Assume, we have a Hilbert space 
$H_{0}$
 and an operator 
$U : H_0 \to H_0$ such that:

\par\addvspace{\medskipamount}\noindent  
{\rm 1)} 
$U$ 
 is linear, bijective, bounded;
\footnote{ hence 
$U^{*}$, 
$U^{-1}$,
$U^{*-1}$
 are bounded }
\par\addvspace{\smallskipamount}\noindent 
{\rm 2)}
$
 S_+(c^{-1}U)=S_+(c^{-1}U^{*-1})=S_+(c^{-1}U^{-1})=S_+(c^{-1}U^*)=\{0\} 
 \,.
$
\par\addvspace{\medskipamount}\noindent 

 Given such an 
$U$,
 let
${\cal U} := {\hat U} = U\oplus U^{*-1}$.
 Then:
\par\addvspace{\medskipamount}\noindent  
 {\rm a)}\quad 
$
 S_+(c^{-1}{\cal U}) = S_+(c^{-1}{\cal U}^{-1}) = \{0\}\,
$
 \hfill\makebox[35ex][l]{ 
 ( by Remark {\bf R1-2} );} 
\par\addvspace{\smallskipamount}\noindent  
 {\rm b)}\quad
$\cal U$ 
 is 
$J$-unitary and symplectic 
 \hfill\makebox[35ex][l]{ 
 ( by Lemma {\bf H1-1});}
\par\addvspace{\smallskipamount}\noindent  
 {\rm c)}\quad 
 items (i), (ii), (iii) are fulfilled 
 \hfill\makebox[35ex][l]{ 
 ( by Remark {\bf R1-1}).}
\par\addvspace{\medskipamount}\noindent  

 Now we have to construct that 
$U$.
 We will do it in Lemma {\bf L2-1},
 but before starting we must introduce some definitions and facts
 related to the theory of so-called weighted shifts.

\begin{Definition}{ D2-1}
 Let 
$H_0$
 be any 
 separable (real or complex) Hilbert space.

 Let 
$(\,,\,)$ 
 stand for the scalar product in
$H_0$
 and let 
$\{b_n\}_n$
 denote an orthonomal basis of 
$H_0$,
 the elements of which are indexed by 
$n=...,-1,0,1,...$.

 Let 
$\{u_n\}_{n\in {\bf Z}}$ 
 be a bilateral sequence; we will suppose that 
$u_n \not= 0$
 for all 
$n\in {\bf Z}$.

 Now let 
$U$ 
 denote the {\bf shift } 
\footnote{ 
 the full name is: the {\bf bilateral weighted shift of 
$\{b_n\}_n$,
 to the right}.
} 
 that is generated by the formula 
$$
 U\,:\,b_n\,\mapsto\,\frac{u_{n+1}}{u_n}\,b_{n+1} \qquad . \eqno(*)
$$
\end{Definition}

 The general facts we need are these:

\begin{Observation}{ O2-1}

 One constructs the 
$U$ as follows:
 
 One starts extending the instruction
$(*)$
 on the linear span of the 
$\{b_n\}_{n\in {\bf Z}}$ 
 so that the resulted operator becomes linear.
 That extension is unique and defines a linear densely defined 
 operator, which is here denoted by
$U_{min}$,
 and which is closable.
 The closure of
$U_{min}$
 is just the 
$U$.

 Now then, this 
$U$
 is closed and at least densely defined and injective;
 it has dense range and the action of
$U^N$, 
$U^{*-N}$, 
$U^{*N}U^N$, 
$U^{-N}U^{*-N}$ 
 (for any integer $N$) is generated by 
$$
 U^N:b_n\mapsto \frac{u_{n+N}}{u_n}b_{n+N}\,;\qquad 
 U^{*-N}:b_n\mapsto \frac{u_n^*}{u_{n+N}^*}b_{n+N}\,;
$$
$$
 U^{*N}U^N:b_n\mapsto {|\frac{u_{n+N}}{u_n}|}^2 b_n\,;\qquad 
 U^{-N}U^{*-N}:b_n\mapsto {|\frac{u_n}{u_{n+N}}|}^2 b_n\,.
$$

 In particular,
$U^N$
 is bounded just when the number sequence 
$\{|u_{n+N}/u_n|\}_n$
 is bounded.
$\Box$
\end{Observation}

 The special factors we need are these: 

\begin{Observation}{ O2-2}
 The family 
$\{b_n\}_n$ 
 is an orthonormal basis. In addition 
$U^Nb_n\perp U^Nb_m$ for 
$n\neq m$. Thus 
$$
 {\|U^Nf\|}^2=\sum_n|(b_n,f)|^2\|U^Nb_n\|^2=
 \sum_n|(b_n,f)|^2{|\frac{u_{n+N}}{u_n}|}^2\,
$$
 for every 
$f\in D_{U^N}$.

 In particular, 
$$\|U^Nf\|\geq|(b_n,f)||u_{n+N}/u_n|$$ 
 for all integers 
$n$. 

 It follows that: 

 Given 
$f\in H_0\setminus \{0\}$
 and given some real $M,a$ such that 
$$
 \|U^Nf\|\leq Ma^N 
 \mbox{ for } 
 N=0,1,2,...\,,
$$
 then there exists a real 
$M'$
 such that 
$$
 |u_N|\leq M'a^N 
 \mbox{ for } 
 N=0,1,2,...
$$
 For
$U^{*-1},U^{-1},U^*$, we have the similar implications. 
 Stated explicitly, they are:
$$
\begin{array}{llcccccccccc}
 
 &
 \|U^Nf\|            &    \leq      &   Ma^N\,        & \Rightarrow &&
 \,    |u_N|         &    \leq      &   M^{'}a^{N}    & 
 \,  (N=0,1,2,...)   &  \\

 &
 \|U^{*-N}f\|        &     \leq     &   Ma^N\,        & \Rightarrow &&
 \,    |u_N|^{-1}    &     \leq     &   M^{'}a^{N}    &
 \,  (N=0,1,2,...)   &  \\

 &
 \|U^{-N}f\|         &     \leq     &    Ma^N\,       & \Rightarrow &&
 \,    |u_{-N}|      &     \leq     &    M^{'}a^{N}   &
 \,  (N=0,1,2,...)   &   \\

 &
 \|U^{*N}f\|         &     \leq     &    Ma^N\,       & \Rightarrow &&
 \,    |u_{-N}|^{-1} &     \leq     &    M^{'}a^{N}   &
 \,    (N=0,1,2,...) &     \\
\end{array}
$$
\footnote{
 We have here meant the format
$$
\begin{array}{cccccc}
 \exists f \in H_0\setminus \{0\}, M>0 , a>0 
 \forall N \geq 0 
 &\cdots 
 & \Rightarrow &
 \exists M'>0 \forall N \geq 0 
 & \cdots 
\end{array}
 \,.
$$
}
 By proving the next Lemma, we will apply exactly such consequences  
 of these implications:

 Let 
$c$
 be a real number such that
$c>0$.
 Then 

$$
\begin{array}{lccccccc}

S_+(c^{-1}U) \not= \{0\}        & \Rightarrow & \exists M' > 0 \forall N \geq 0
& \,    |u_N|          &  \leq  &   M^{'}(c+1)^{N}    & 
\\

S_+(c^{-1}U^{*-1}) \not= \{0\}  & \Rightarrow & \exists M' > 0 \forall N \geq 0
& \,   |u_N|^{-1}      &   \leq &   M^{'}(c+1)^{N}    &
\\

S_+(c^{-1}U^{-1}) \not= \{0\}   & \Rightarrow & \exists M' > 0 \forall N \geq 0
& \,   |u_{-N}|        &   \leq &    M^{'}(c+1)^{N}   &
\\

S_+(c^{-1}U^*) \not= \{0\}      & \Rightarrow & \exists M' > 0 \forall N \geq 0
& \,   |u_{-N}|^{-1}   &   \leq &    M^{'}(c+1)^{N}   &
\\

\end{array}
$$
$\Box $
\end{Observation}

\begin{Lemma}{ L2-1}
 Let 
$c$
 be a real number such that
$c>0$.

 Set 
$$
 u_n\,:=\,(c+2)^{|n|\sin(\frac{\pi}{2}\log_2(1+|n|))} 
 \qquad (n=...,-1,0,1,...)
$$
 
 Then the associated shift 
$U$
 is bounded with its inverse and
$$
 S_+(c^{-1}U) = S_+(c^{-1}U^{*-1}) = S_+(c^{-1}U^{-1})
 = S_+(c^{-1}U^*) = \{0\} \,.
$$
\end{Lemma}

\begin{Proof}{ }
 The derivative of the real-valued function
$$
 x\,\mapsto\,|x|\sin(\frac{\pi}{2}\log_2(1+|x|))
$$
 is equal to 
$$
 (\sin(\frac{\pi}{2}\log_2(1+|x|))+\frac{\pi}{2\ln 2}\frac{|x|}{1+|x|}
 \cos(\frac{\pi}{2}\log_2(1+|x|))){\rm sgn}\,x
$$
 and its absolute value does not exceed the value of 
$\alpha:=1+\pi/(2\ln 2)$.
 By the Mean Value Theorem ( Lagrange ), 
$$
 (c+2)^{-\alpha}\,\leq\,|u_{n+1}/u_n|\,\leq\,(c+2)^{\alpha} \,.
$$
 Hence 
$U$
 and 
$U^{-1}$
 are bounded.

 Now choose two sequences of integers defining them by 
$$
 n_k\,:=\,2^{1+4k}-1;\quad m_k\,:=2^{3+4k}-1 \qquad (k=1,2,...) \,.
$$
 Then 
$n_k,\, m_k \in {\bf N}$,\ 
$n_k\,\to\,+\infty$,$m_k\,\to\,+\infty$
 (as 
$k\,\to\,+\infty$),
 and simultaneously 
$$
 u_{n_k} = u_{-n_k} = (c+2)^{n_k} \,; \quad
 u_{m_k}^{-1} = u_{-m_k}^{-1} = (c+2)^{m_k} \,.
$$

\noindent
 We see that no estimation of the form 

$$
\begin{array}{clc}
 |u_N|\,\leq\,M'(c+1)^N,\,
&
 |u_{-N}|\,\leq\,M'(c+1)^N,\,
&
 |u_{N}|^{-1}\,\leq\,M'(c+1)^N,\,
\\
&
 |u_{-N}|^{-1}\,\leq\,M'(c+1)^N
&
\end{array}
$$

\noindent 
 (for 
$N=0,1,\cdots$
 )
 is possible. 
 On looking at the Observation {\bf O2-2}, we see that 
$$
S_+(c^{-1}U)=\{0\} \,,\,
S_+(c^{-1}U^{-1})=\{0\} \,,\,
S_+(c^{-1}U^{*-1})=\{0\} \,,\,
S_+(c^{-1}U^*)=\{0\} \,.\,
$$
 This is just what was to be proven.

\end{Proof}
 The proof of Lemma {\bf L2-1}
 is completed, so is the proof of Theorem {\bf Th2-1}.
\end{Proof}

\begin{Remark}{ R2-1}
 Actually, we have taken a number 
$c>0$ 
 quite arbitrarily, and then we have constructed a system, 
 such that {\bf both} of the Ljapunov indices 
$\lambda_{\pm}$ 
 of {\bf every } non-zero displacement 
$\geq\ln(2+c)$.
 And at the same time, there are displacements such that their 
 {\bf lower } index of growth 
$ = -\ln(2+c)$.
\end{Remark}


\newpage\section
{ Another Examples of 
$J$-unitary Operators }

 In this section, we construct two more operators which properties 
 looks something strange.
 The elements of the constructions are the same as that which 
 we have introduced in the previous sections, namely:

\smallskip 

$H_0$, 
 it stands for any separable Hilbert space;
$\{b_n\}_n$,
 it stands for a orthonormal basis of 
$H_0$,
 the elements of that basis will be indexed by 
$n=...,-1,0,1,...$ .
 In addition, 
$$
 \hat H_0:=H_0\oplus H_0 \,,\quad 
 J(x\oplus y):=y\oplus x \,,\quad 
 {\cal J}(x\oplus y):= -y\oplus x \,,\quad (x,y \in H_0) 
$$ 
 and given a linear operator
$ T: H_0 \to H_0$ , 
 we put 
$ \hat T:= T\oplus T^{* -1}$, 
 whenever 
$T^{* -1}$
 exists. 

\smallskip 
 
 We construct two bilateral sequences of numbers 
$\{v_n\}_n$, 
$\{w_n\}_n$, 
$n=...,-1,0,1,...$, 
 so that the associated shifts, 
$V$
 and 
$W$, 
 and the corresponding 
$J$-unitary and symplectic automorphisms, 
$\hat V$
 and 
$\hat W$, 
 have especial properties.

\begin{Definition}{ D3-1} 
 Let 
$c$
 be a real number such that
$c\geq 1$.
 Let 
$v_n:= (2c)^{-|n|}$ 
 for any integer 
$n$.
 Let 
$V :H_0 \to H_0$ 
 denote the associated shift, defined by 
$$
 V\, : \, b_n \mapsto \frac{v_{n+1}}{v_n} b_{n+1} \,.
$$ 
 With other words, let 
$$
 Vb_n :=\frac{1}{2c} b_{n+1} \, \hbox{ for } n=0,1,2,... \quad
 Vb_n := 2cb_{n+1} \, \hbox{ for } \quad n=...,-2,-1.
$$ 
\end{Definition}

\begin{Remark}{ R3-1}
 The just now defined 
$V$
 is bounded and invertible and its inverse is bounded as well.
 Using the definition one can show that 
\par\addvspace{\smallskipamount}\noindent 
 {\rm (1)} \quad 
$
 \| V^N b_n \| = (2c)^{-|n+N|+|n|} \quad
  \mbox{ for all integers } \quad n, N;
$ 
\par\addvspace{\smallskipamount}\noindent 
 {\rm (2)} \quad 

$ r(V) = r (V^{-1}) = 2c;$
\par\addvspace{\smallskipamount}\noindent 
 {\rm (3)} \quad  
$
 \overline{ S_0 \left(\frac{3c}{2} V \right)} = H_0, \,
   S_0 \left(\frac{2}{3c} V^{*-1} \right) = \{0\}, \,
 \overline{ S_0 \left(\frac{3c}{2} V^{-1} \right)} = H_0, \,
   S_0 \left(\frac{2}{3c} V^{*} \right) = \{0\}.
$ 
\\
\end{Remark}

\begin{Lemma}{ L3-1}
 Let
$L$, 
$M$ 
 be (linear closed) subspaces of 
$\hat H_0$ 
 such that 
$$
 \hat V L = L , 
 \quad | spectrum\, \hat V | L | \, \leq \, c \, ,
 \hat V^{-1} M = M , 
 \quad | spectrum\, \hat V^{-1} | M | \, \leq \, c \, .
$$
 Then: 
\par\addvspace{\smallskipamount}\noindent 
{\rm (a)}\hfill 
\parbox[t]{.93\textwidth}{ 
$ L = L_1 \oplus \{ 0\}, \, \quad
  M = M_1 \oplus \{ 0\}, \,
  \mbox{ for some } \,
    L_1 \, \subset \, H_0 \,,\, 
    M_1 \, \subset \, H_0 ;
$
}
\par\addvspace{\smallskipamount}\noindent 
{\rm (b)}\hfill 
\parbox[t]{.93\textwidth}{ 
$ VL_1 = L_1$, 
$| spectrum\, V | L_1 | \, \leq \, c;$
\\
$ V^{-1}M_1 = M_1$,
$| spectrum\, V^{-1} | M_1 | \, \leq \, c;$
}
\par\addvspace{\smallskipamount}\noindent
{\rm (c)}\hfill 
\parbox[t]{.93\textwidth}{
$  L_1 \, \not= \, H_0 ,\quad  M_1 \, \not= \, H_0$ .
}
\end{Lemma}

\begin{Proof}{ }

 Proof of (a) :
 Follow from 
\begin{eqnarray*} 
  L \, \subset \, S_0 \left( \frac{2}{3c} \hat V \right)
&=&
  S_0 \left( \frac{2}{3c} V \oplus \frac{2}{3c} V^{*-1} \right)
\\
&=&
  S_0 \left( \frac{2}{3c} V \right) \,
   \oplus \,
     S_0 \left( \frac{2}{3c} V^{*-1} \right) 
   =
     S_0 \left( \frac{2}{3c} V \right) \, \oplus \, \{ 0 \};
\\
 M \, \subset \, S_0 \left( \frac{2}{3c} \hat V^{-1} \right)
&=&
  S_0 \left( \frac{2}{3c} V^{-1} \oplus \frac{2}{3c} V^{*} \right)
\\
&=&
  S_0 \left( \frac{2}{3c} V^{-1} \right) \,
   \oplus \, 
     S_0 \left( \frac{2}{3c} V^{*} \right)
   =
     S_0 \left( \frac{2}{3c} V^{-1} \right) \, \oplus \, \{ 0 \};
\end{eqnarray*}

 Proof of (b): 
 After (a) is proven, we can state:
$$
 L = L_1 \oplus \{ 0 \}, \, M = M_1 \oplus \{ 0 \}, \, \hat V=V\oplus V^{*-1}.
$$
 Hence 
$$ 
 spectrum\,\hat V|L = spectrum\,V|L_1, \quad 
 spectrum\,\hat V^{-1}|M = spectrum\, V^{-1}|M_1. 
$$
 Therefore 
$$ |spectrum\,V|L_1| =| spectrum\,\hat V|L| \leq c, $$ 
$$ |spectrum\,V^{-1}|M_1| = |spectrum\,\hat V^{-1}|M| \leq c. $$

 Proof of (c) : 
 We have 
\\
$ | spectrum\, V | L_1 | \, \leq \, c$,
 and 
$r(V)=2c$.
 Hence 
$ L_1 \not= H_0 $ .
 Similarly, 
$ | spectrum\, V^{-1} | M_1 | \, \leq \, c$,
 and 
$r(V^{-1})=2c$ . 
 Hence 
$ M_1 \not= H_0 $ . 

\end{Proof}

 Now recall that 
$H\oplus \{0\}$
 is 
$J$-neutral subspace of 
$\hat H_0$
 (see [Krein65]). 
 In particular, 
$H_0 \oplus \{0\}$
 is a semidefinite subspace.

 So, we now come to 

\begin{Theorem}{ Th3-1}
 Let 
$L$
 be a semidefinite subspace of 
$\hat H_0$ 
 such that 
$$
\hat V L = L, \qquad |spectrum\, \hat V|L | \, \leq \, c .
$$
 Then 
$L$
 is not maximal.

 Let 
$M$
 be a semidefinite subspace of 
$\hat H_0$ 
 such that 
$$
\hat V^{-1} M = M, \qquad |spectrum\, \hat V^{-1}|M | \, \leq \, c .
$$
 Then 
$M$
 is not maximal.

\end{Theorem}

\begin{Remark}{ R3-2}

$V$
 has an interesting property:

$b_n \in S_0(V) \cap S_0(V^{-1})$ 
 for every integer 
$n$;
 as a result, 
$S_0(V) \cap S_0(V^{-1})$
 is dense in 
$H_0$ .
 Moreover, let 
$L_k$ denote the closed linaer span of
$\{V^s b_k|\, s\geq k\}$. 
 Then 
$$
 VL_k\,\subset L_k \ ,
 r(V|L_k) \,\le\,1/2c \ 
 \mbox{ and } 
 H_0 = \overline{\cup\{L_k\,|k=...-1,0,1...\} }\ .
$$
 In spite of that 
$r(V)=2c$ .

$V^{-1}$ has the similar property.
 But there 
$L_k$ is to be replaced by the closed linaer span of 
$\{V^{-s}b_k|\,s\geq k \}$ .

\addvspace{\bigskipamount}

 What we now want to know is what kind of growth of 
$ {\|V^Nf\|}^2 $
 is possible,
 as 
$N \to \pm \infty $ .
 Is there an
$f \in H_0$
 such that
$ {\|V^Nf\|}^2 \to \infty $, 
 as 
$N \to \pm \infty $ ?

\addvspace{\bigskipamount}
 
 Let 
$f:= {\sum}_{n\not=0}|n|^{-1} b_n$. 
 Then 
$f \in H_0$
 and 
\begin{eqnarray*}
 {\|V^Nf\|}^2
 &=& \sum_n|(b_n,f)|^2{|\frac{v_{n+N}}{v_n}|}^2\,
\\
 &=&\sum_{n>0}\frac{1}{|n|^{2}}\frac{(2c)^{-2|n+N|}}{(2c)^{-2|n|}}\,
    +\sum_{n<0}\frac{1}{|n|^{2}}\frac{(2c)^{-2|n+N|}}{(2c)^{-2|n|}}\,
\\
 &=&\sum_{n>0}\frac{1}{|n|^{2}}\frac{(2c)^{-2|n+N|}}{(2c)^{-2|n|}}\,
    +\sum_{n>0}\frac{1}{|n|^{2}}\frac{(2c)^{-2|n-N|}}{(2c)^{-2|n|}}\,
\end{eqnarray*}

 Let us estimate the sequence 
$ {\|V^Nf\|}^2 $ 
 {\bf from below }.
 
 Let 
$N\geq 0$. 
 Then 
\begin{eqnarray*}
 {\|V^Nf\|}^2 
 &\geq&
    +\sum_{n>N}\frac{1}{|n|^{2}}\frac{(2c)^{-2|n-N|}}{(2c)^{-2|n|}} 
\\
 &&
    =\sum_{n>N}\frac{1}{|n|^{2}}\frac{(2c)^{-2n+2N}}{(2c)^{-2n}} 
\\
 &&
    =\sum_{n>N}\frac{1}{|n|^{2}}(2c)^{2N} 
\\
 &&
    \geq\frac{1}{N+1}(2c)^{2N}
    \geq\frac{1}{2^N}(2c)^{2N}
    = 2^N c^{2N} =2^{|N|} c^{2|N|} \,.
\end{eqnarray*}

 Besides we observe that for the current 
$V$
 and 
$f$ 
 the quantity 
${\|V^Nf\|}^2$
 depends on 
$N$ 
 so that 
${\|V^Nf\|}^2 = {\|V^{-N}f\|}^2$ 
 for all 
$N$.

 Thus we have seen that 
$$ 
 {\|V^Nf\|}^2 
    \geq\frac{1}{|N|+1}(2c)^{2N}
       \geq 2^{|N|} c^{2|N|} \mbox{ for all } N \,. 
$$ 
 A very rapid growth of 
$ {\|V^Nf\|}^2 $, 
 as 
$N \to \pm \infty $
 !!

\end{Remark}

 The example of 
$J$-unitary and symplectic automorphism we are now describing shows 
 that two mathematecally very natural formulations of the phrase 
 `` ... is stable with respect to the action of ... ''
 can in the real situation appear as ``orthogonal'' to one another.

\begin{Definition}{ D3-4}
 Let 
$w_n:= 2^{-|n|} =2^n$
 for 
$n \leq 0$
 and 
$ w_n:= 1/(n+1)$
 for 
$n > 0$.
 Let 
$W :H_0 \to H_0$ 
 denote the associated shift generated by 
$$
 W\, : \, b_n \mapsto \frac{w_{n+1}}{w_n} b_{n+1} \,.
$$ 
 
\end{Definition}

\begin{Remark}{ R3-3}
 Since 
$1/2 \leq w_{n+1}/w_n \leq 2$ 
 for all integers 
$n$, $W$
 is bounded invertible 
 and 
$W^{-1}$
 is bounded as well.
\end{Remark}

\begin{Lemma}{ L3-2}
 That just now defined $W$ has the properties:

$$
 \overline{S_0 (W)} = H_0, \quad 1 \leq |spectrum\, W | \leq 2, \quad r(W)= 2,
$$
$$
 S (W^{*-1}) = \{ 0\}, \quad \frac12 \leq |spectrum\, W^{*-1} | \leq 1, 
\quad r (W^{*-1}) = 1 \,.
$$

\end{Lemma}

\begin{Proof}{. }
 The proof is founded on the well-known formula for spectral radius,
 on Remark {\bf R3-3} and on the formulae in Observation {\bf O2-1}. 
 We have: 
$$
 \|W^N\| = sup \{ \frac{w_{n+N}}{w_n} \, | \, n= \dots -1, 0, 1, \dots \}, 
$$
$$
 \| W^{*-N} \| = sup \{ \frac{w_n}{w_{n+N}} \, | \, 
 n= \dots -1, 0, 1, \dots \} .
$$

 Take 
$N>0$
 arbitrarily, and analyse the 
$w_n /w_{n+N}$
 in details.
 We observe: 


\begin{description}
\item[\rm a)]
$ w_n /w_{n+N} = 1/2^N $
 for 
$n +N \leq 0$;\\
\item[\rm b)]
$ w_n /w_{n+N} = (1+n+N)/(1+n) = N/(n+1)\, + \, 1 
\, \leq \, N+1$
 \hspace*{7mm} for 
$0 < n$;\\
\item[\rm c)]
$ w_n /w_{n+N} = 2^n (1+n+N)\, \leq \, N+1$
 for 
$n\leq 0 < N+n$
\end{description}

 Therefore 
$\|W^{*-N}\|\, \leq N+1$
 (for 
$N>0$) .
 Note that 
$w_0/w_N = 1+N$ .
 Hence 
$\|W^{*-N}\|\,=N+1$
 (for 
$N>0$) .
 Therefore 
$r(W^{*-1}) = 1$
 and 
$ r(W^{-1})=1$ .
 Quite similarly we can analyse 
$r(W)$ 
 and 
$r(W^*)$ :
 Note 
$\|W^N\| = 2^N$ (for $N>0$). 
 Therefore 
$r(W) = 2$ and $r(W^*) = 2$.

 Finally, if 
$n+N > 0$,
 then 
$W^N b_n = {w_n}^{-1}(1+N+n)^{-1}b_{n+N}$.
 Therefore 
$b_n\in S_0(W)$
 for all integers 
$n$.
 Hence 
$\overline{S_0(W)} = H$
 and 
\mbox{$ S(W^{*-1})=\{0\}\,. $}

\end{Proof}

\begin{Theorem}{ Th3-2}

 There exists a $J$-unitary operator
$\hat W$ 
 and a maximal semidefinite subspace 
$L$
 such that: 
\par\addvspace{\bigskipamount}\noindent 
{\rm (a)}\hfill 
\parbox[t]{.93\textwidth}{ 
 $\hat WL^\perp = L^\perp 
 \,,\quad 1\leq |spectrum\hat W|L^\perp \,| \, \leq \, 2,\quad
   r(\hat W|L^\perp )\,=2 $ 
 \medskip \\ 
 but in spite of that, 
$L^\perp = \overline{S_0(\hat W)} \,=\, \overline{S(\hat W)}$.\\
 }
\par\addvspace{\smallskipamount}\noindent 
{\rm (b)}\hfill 
\parbox[t]{.93\textwidth}{ 
$ \hat WL = L ,\quad |spectrum\hat W|L | \, \leq \, 1 $,
 \medskip \\ 
 although 
$ L \cap \overline{ S(\hat W )} = \{0\}$. 
 }

\end{Theorem}

\begin{Proof}{ }
 Set 
$$
 L:=\{0\}\oplus H_0 \,,
 \, M:=H_0\oplus \{0\} \equiv L^\perp \,,
$$ 
 and apply Lemma {\bf L3-2} to the formulae for 
$S_0(\hat W)$
 and 
$S(\hat W)$
 (see Introduction):
$$
 S_0(\hat W)\,=S_0(W)\oplus S_0(W^{*-1})\,=S_0(W)\,\oplus \{0\}\,\subset M 
$$
$$
 S(\hat W)\,=S(W)\oplus S(W^{*-1})\,=S(W)\,\oplus \{0\}\,\subset M
$$
 But 
$S_0(W)$
 and 
$S(W)$,
 both of them are dense in 
$H_0$.
 Hence the closures of 
$S_0(\hat W)$
 and 
$S(\hat W)$
 coincide with
$M$.
 To complete the proof, note that 
$\hat W|M $ 
 is unitarily equivalent to 
$W$, 
$\hat W|L $ 
 is unitarily equivalent to 
$W^{*-1}$,
 and then again apply Lemma {\bf L3-2}.
\end{Proof}


\newpage\section
{ Coming to Models of Dynamics in Continuous Time }


 A quite traditional way to obtain 
 a model of dynamics in continuous time 
 from 
 a given model of dynamics in discrete time
 consists in rewriting the relations of the latter replacing,
 in appropriate positions, 
 symbols of sequences (functions of a discrete time)
 by symbols of functions of a continuous time 
\footnote{ symbols that suggest that
 ``this object is a function defined on a set having a discrete structure''
 by symbols that suggest that
 ``this object is a function defined on a set having a continuous structure''
 } 
, 
 symbols of discrete-valued (integer-valued) variables representing time,
 by symbols of continuum-valued (real-valued) variables,
 that remain to call ``time'',
 provided by a suitable redefining such notions as ``sum'', and all that.

 So, in this way, the definition of the shift given in a previous 
 section is being transformed as follows:

$$
 V^N: b_n \mapsto \frac{v_{n+N}}{v_n} b_{n+N}
$$
$$
 V^N: \sum_n f(n) b_n \mapsto \sum_n \frac{v_{n+N}}{v_n}f(n)b_{n+N}
                            =\sum_n \frac{v_n}{v_{n-N}}f(n-N)b_n\,;
$$
$$
 V^N: f(n) \mapsto \frac{v_n}{v_{n-N}}f(n-N)
$$
$$
 V(t): f(x) \mapsto \frac{v(x)}{v(x-t)}f(x-t)
$$
 Notice 
$$
\begin{array}{rcr}
 V(t)V(\tau )^{-1} \, : 
 &&
 \\
 f(x) 
 & 
 \stackrel{V(\tau )^{-1}}{\longmapsto}\dfrac{v(x)}{v(x+\tau)}f(x+\tau)
      \stackrel{V(t)}{\longmapsto}
 & \dfrac{v(x)}{v(x-t)}\dfrac{v(x-t)}{v(x-t+\tau)}f(x-t+\tau) %
 \\&&{} =\left( V(t-\tau)f \right)(x) 
\end{array}
$$
 With other words the dynamics generated by 
$V$
 is time-autonomous and its formal generator is:
\begin{eqnarray*}
\left(Hf\right)(x)&=&\left(V(t)f\right)_{t=0}'(x)  \\
&=&\frac{\partial}{\partial t}\left[\frac{v(x)}{v(x-t)}f(x-t)\right] _{t=0} \\
&=&-\frac{\partial f(x)}{\partial x} +\frac{v'(x)}{v(x)}f(x)
  = -v(x)\frac{\partial }{\partial x}\Bigl(\frac{1}{v(x)}f(x)\Bigr)  \,.
\end{eqnarray*}

 If we apply the conversion method presented above
 especially to the discrete systems,   
 which we described in the previous sections,
 we will see that the corresponding continuous systems 
 may be described as:
\begin{description}
\item[\rm a)]
$v(x)=e^{|x|sin\big(ln(1+|x|)\big)}$ \\
$(Hf)(x)$  \\
${ }=-\dfrac{\partial f(x)}{\partial x}
 +\Big(sin(ln(1+|x|))
  +\dfrac{|x|}{1+|x|}cos\bigl(ln(1+|x|)\bigr)\Big)sgn(x)f(x) $
\\
\item[\rm b)]
$ v(x)=e^{-|x|}$\\
$(Hf)(x)=-\dfrac{\partial f(x)}{\partial x}-(sgn\,x)f(x)$\\
\item[\rm c)]
$$
 v(x)=\left\{\begin{array}{lcl} 
 e^x&,&x<0\\
\dfrac{1}{x+1}&,&x>0 
\end{array} \right\}
$$
$$
 (Hf)(x)=-\frac{\partial f(x)}{\partial x}
 +\left\{
 {1 ,x<0 \atop -\dfrac{1}{x+1} ,x>0 }
 \right\}
 f(x) 
$$
\end{description}
 
 Of course, the behaviour of these systems is irregular likewise 
 the the behaviour of their prototypes; but we will not here discuss it.

\newpage\section
{ Appendix A. } 
\subsection* 
{ A Possible Definition of Abstract Linear Dynamical System } 
 
\begin{Definition}%
{ of abstract linear dynamical system.}

\noindent
 Let 
$L$
 be a linear space,
${\cal T}_{\cal A}$
 be a set (abstract time),
$\geq$
 be a transitive relation on 
${\cal T}_{\cal A}$.

\noindent
 Let a two-parameter family of linear operators on 
$L$,

$$
 \{V_{t,s}\}_{t,s}  \qquad  ( t\geq s \,\quad t,s \in {\cal T}_{\cal A} ) \,,
$$
 be such that 
$$
  V_{t,r}V_{r,s} = V_{t,s} \,, \qquad \mbox{\rm if } t\geq r\geq s \qquad  
  \makebox[12ex][l]{ {\bf ( consistency relation ) } }
$$

\par\addvspace{\medskipamount}\noindent
 In this case the structure  
$$
 L \quad ,\quad
 {\cal T}_{\cal A} \quad ,\quad 
  \geq \quad ,\quad
 \{V_{t,s}\}_{t,s}
$$
 is said to be an {\bf abstract linear dynamical system}
 and 
$ \{V_{t,s}\}_{t,s} $
 is called a {\bf propagator}, alias {\bf evolution operator}.
%
\footnote{ we often write 
$V_{t,s}$
 instead of 
$ \{V_{t,s}\}_{t,s} $
 } 
.

\par\addvspace{\medskipamount}\noindent
 Given a 
$ t_0 \in {\cal T}_{\cal A} $
 and a 
$x_0 \in L$,
 we say that the one-parameter family,
$\{x(t)\}_{ t \geq t_0 }$,
 defined by 
$$
 x(t) =  V_{t,t_0}x_0  \qquad (t\geq t_0) \,,
$$ 
 is a (future or forward) {\bf trajectory}. In such a case we say that 
$t_0, x_0$
 are {\bf initial data}.

\par\addvspace{\medskipamount}\noindent

 If the propagator is such that each 
$
 V_{t,s} 
$
 is invertible, then we say that the dynamics is {\bf invertible}.
 In this case we put 
$$
 V_{s,t} := V_{t,s}^{-1}  \qquad (t\geq s) \,.
$$

  Finally, if 
${\cal T}_{\cal A}$
 is equipped with the discrete topology,
 we say that the dynamics is {\bf discrete}.
\end{Definition}






\newpage
\bibliographystyle{unsrt}

\end{document}